\documentclass[conference]{IEEEtran}
\ifCLASSINFOpdf
\else
\fi
\usepackage{graphicx}
\usepackage{float}
\usepackage{subfig}
\usepackage{CJK}
\usepackage{amsmath}
\usepackage{amsthm}

\usepackage[justification=centering]{caption}
\begin{document}
\title{A Workload-Specific Memory Capacity Configuration Approach for In-Memory Data Analytic Platforms}

\author{\IEEEauthorblockN{Yi~Liang\\}
\IEEEauthorblockA{Beijing University of \\Technology, Beijing, China\\
Email: yliang@bjut.edu.cn}
\and
\IEEEauthorblockN{Shilu~Chang\\}
\IEEEauthorblockA{Beijing University of \\Technology, Beijing, China\\
Email: bryant\_chang@emails.bjut.edu.cn}
\and
\IEEEauthorblockN{Chao~Su\\}
\IEEEauthorblockA{Beijing University of \\Technology, Beijing, China\\
Email: S201607123@emails.bjut.edu.cn}}% <-this % stops a space
% \thanks{M. Shell was with the Department
% of Electrical and Computer Engineering, Georgia Institute of Technology, Atlanta,
% GA, 30332 USA e-mail: (see http://www.michaelshell.org/contact.html).}% <-this % stops a space
% \thanks{J. Doe and J. Doe are with Anonymous University.}% <-this % stops a space
% \thanks{Manuscript received April 19, 2005; revised August 26, 2015.}

% \markboth{Journal of \LaTeX\ Class Files,~Vol.~14, No.~8, August~2015}%
% {Shell \MakeLowercase{\textit{et al.}}: Bare Demo of IEEEtran.cls for IEEE Journals}

% \institute{Beijing University of Technology, Beijing, China}
\maketitle

\begin{abstract}
Nowadays, in-memory data analytic platforms, such as Spark, are widely adopted in big data processing. The proper memory capacity configuration has been proved to be an efficient way to guarantee the workload performance in such platforms.Currently, Spark adopts the static way to configure the memory capacity for workloads based on user specifications. However, due to the lack of deep knowledge of the target platform and workload characteristics, nonexpert users often conservatively configure the memory capacity in an excessive way, which reduces the memory utilization significantly. On the other hand, as the memory requirements are quite different among diverse workloads, there is not the one-size-fits-all solution for memory capacity configuration.Aiming on these issues, we propose WSMC, a workload-specific memory capacity configuration approach for the Spark workloads, which guides users on the memory capacity configuration with the accurate prediction of the workload's memory requirement under various input data size and parameter settings.First, WSMC classifies the in-memory computing workloads into four categories according to the workloads' Data Expansion Ratio. Second, WSMC establishes a memory requirement prediction model with the consideration of the input data size, the shuffle data size, the parallelism of the workloads and the data block size. Finally, for each workload category, WSMC calculates the shuffle data size in the prediction model in a workload-specific way. For the ad-hoc workload, WSMC can profile its Data Expansion Ratio with small-sized input data and decide the category that the workload falls into. Users can then determine the accurate configuration in accordance with the corresponding memory requirement prediction.Through the comprehensive evaluations with SparkBench workloads, we found that, contrasting with the default configuration, configuration with the guide of WSMC can save over 40\% memory capacity with the workload performance slight degradation (only 5\%), and compared to the proper configuration found out manually, the configuration with the guide of WSMC leads to only 7\% increase in the memory waste with the workload's performance slight improvement (about 1\%)
\end{abstract}

% \begin{IEEEkeywords}
% IEEE, IEEEtran, journal, \LaTeX, paper, template.
% \end{IEEEkeywords}

\IEEEpeerreviewmaketitle

\section{Introduction}
\IEEEPARstart
{I}{n-memory} computing frameworks, are widely adopted in big data processing. These frameworks keep the reused data among multiple tasks in the memory, which can reduce data processing time effectively, especially for iterative and real-time workloads.

As one of the representative in-memory frameworks, Spark[2] has the comprehensive ecosystem, which includes SQL query[3], machine learning, graph computing and streaming[4]. The comprehensive ecosystem makes Spark become the most widely used in-memory framework.
In Spark, memory is required for not only the reused data caching but also the data processing and data shuffling. Hence, the proper memory capacity configuration is a dominant factor of the performance guarantee of Spark workloads. Currently, Spark adopts the static way to configure the memory capacity for workloads based on user specifications. However, determing the proper configuration requires users to have deep knowledge about the target platform and their workloads. This is hard and cumbersome due to that users often employ a third-party packaged application and is not clear of its system-level characeristics. Hence, the non-expert users usually conservatively overestimate the memory consumption of their workloads and require in an excessive way. A recent analysis[6] on the resource utilization over 29 days trace period on Google’s big data center revealed that over 80\% of cluster memory are allocated to big data applications at most time, while around 50\% of them are technically unused. It is obvious that the user specification-based way will reduce the resource utilization significantly, especially under the multi-tenancy environment where the memory is shared and competed.

Guiding users based on the knowledge of platform mechanisms and workload characteristics is a natural and promising alternative on the memory configuration. However, this work is challenging due to two reasons. First, the efficient guiding requires the accurate prediction of the memory demand on each execution stage of the workload, which is not available yet. Second, such a {\textquoteleft}proper configuration{\textquoteright} is different across workloads due to the diverse memory consumption characteristics. However,
such consumption characteristics are not identified and classified among Spark workloads yet. To solve these problems, in this paper, for the Spark platform, we propose WSMC, a workloads-specific memory capacity configuration approach. WSMC classifies Spark workloads into four categories according to their memory consumption characteristic. For each category,the workload{\textquoteright}s memory requirement prediction model is established. On configuring the memory capacity of the ad-hoc workload, WSMC match the workload{\textquoteright}s memory consumption characteristic to one of the above four categories and determine the accurate configuration in accordance with the corresponding memory requirement prediction. Specifically, this paper makes following contributions:
\begin{enumerate}
\item
We explore the Data Expansion Ratio as the specific memory consumption characteristic to classify Spark workloads. The Data Expansion Ratio is referred as the maximum ratio of the shuffle data to the input data among the multiple processing stages of a Spark workload.For detail,according to the value range and the increasing rate of Data Expansion Ratio, Spark workloads can be classified into four categories, including Expanding.Rapid, Expanding.Medium, Medium and Shrinking. For an individual ad-hoc workload, its Data Expansion Ratio can be figured out by executing it with a small set of input data, and then, the corresponding category to be matched can be determined.
\item
We establish a workload-specific memory requirement prediction model for Spark workloads with the parameters of the input data size, the shuffle data size, the task parallelism and the data block size. Especially, we take the shuffle data size as the expansion of the input data size and determine the expansion factor for each workload category respectively. This model can predict the maximum amount of memory required among the multiple stages of Spark workload, and take it as the memory capacity configuration decision.
\item
We evaluate WSMC with SparkBench[7] workloads.For each workload, we execute it with three memory capacity configurations: the default configuration by Spark, the proper configuration found manually, and the configuration guided with WSMC. Compare to the default configuration, the configuration with the guide of WSMC can help to save over 40\% of memory capacity with the workload’s performance slight degradation (only 5\%). Compared to the proper configuration, the configuration with the guide of WSMC leads to only 7\% increase in the memory waste with the workload's performance slight improvement (about 1\%).
\end{enumerate}
\par
The rest of the paper is organized as follows. In Section 2, we discuss the background and our motivation. Section 3 introduces the methodology of workload-specific memory configuration. Section 4 is the evaluation. Section 5 shows the related work. Finally, we draw the conclusion in Section 6.
\begin{figure}[!t]
\centerline{\includegraphics[width=2in]{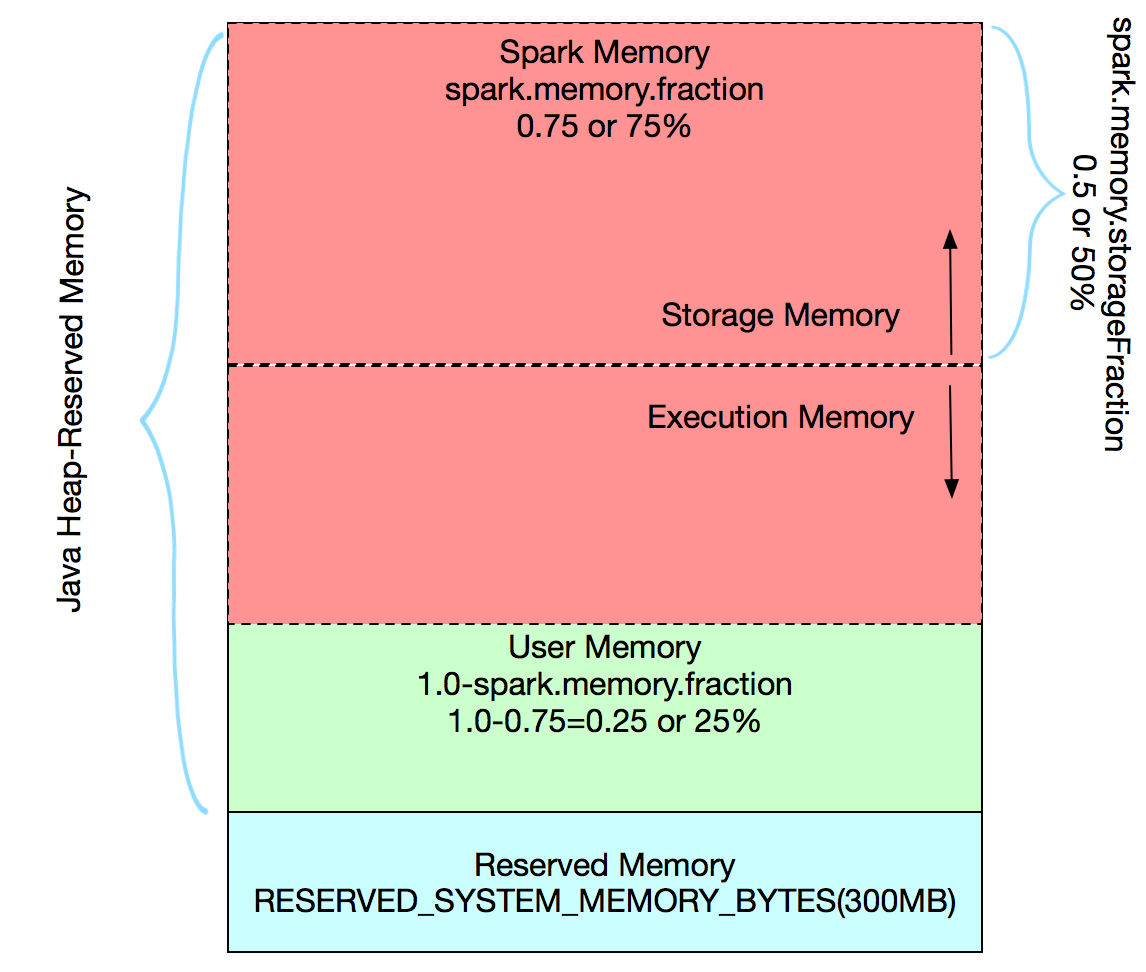}}
\caption{Spark Memory Management Model[1]}
\label{fig_sim}
\end{figure}

\begin{figure*}[!t]
\centering
\subfloat[PageRank]{\includegraphics[width=1.7in]{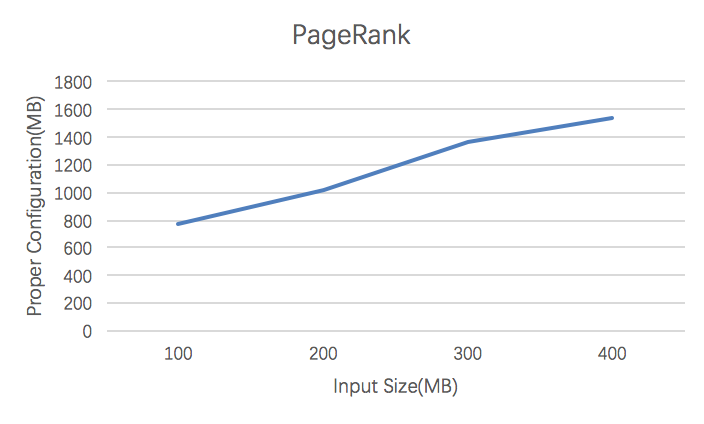}%
\label{fig_first_case}}
\hfil
\subfloat[ShortestPath]{\includegraphics[width=1.7in]{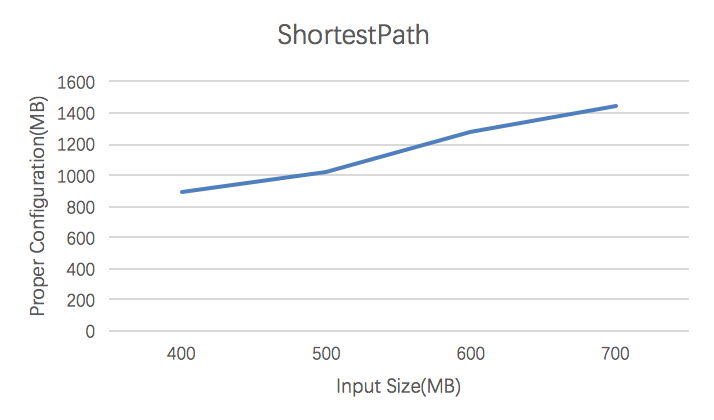}%
\label{fig_second_case}}
\hfil
\subfloat[Terasort]{\includegraphics[width=1.7in]{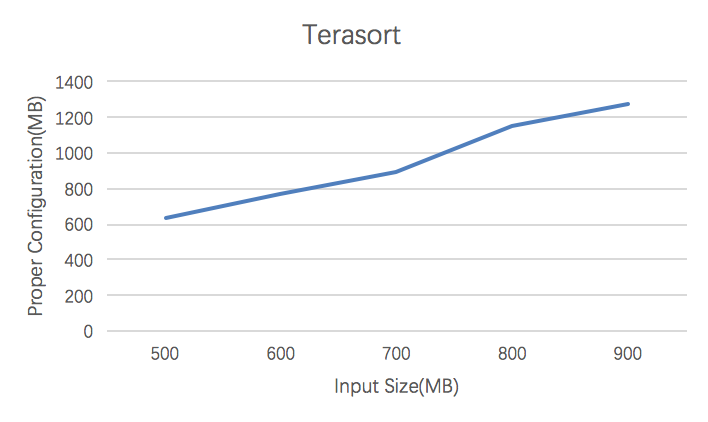}%
\label{fig_first_case}}
\hfil
\subfloat[ConnectComponent]{\includegraphics[width=1.7in]{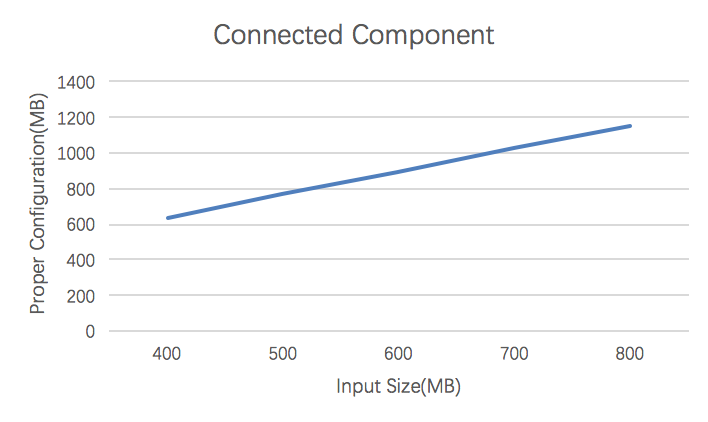}%
\label{fig_first_case}}
\caption{Memory capacity requirements under different input data size}
\label{fig_sim}
\end{figure*}

\section{Background and Motivation}
In this section, we first discuss the memory management mechanism of Spark. Then, we describe our motivation by empirical observations of Spark workloads.
\subsection{Spark Memory Management}
Data is managed as memory abstractions called resilient distributed datasets (RDDs)[8] in Spark platform. The RDD is a collection of objects partitions across a set of nodes, and each node retains several blocks. One RDD block can be replicated across nodes for high availability, and can be recomputed according to the dependency of RDD. If the data lost due to the machine failure, computation is conducted in form of RDD actions and transformations, which can be used to organized the dependency of dataset as a DAG of RDDs.

The Spark workload is deployed as a \emph{driver program} running on a master node of a resource cluster and several \emph{executors} running on worker nodes. Executor launched as a JAVA processes where all tasks are executed in. Fig. 1 shows the typical memory model of the latest version of Spark. Generally, the memory of Spark executor consists of three parts: \emph{Reserved Memory}, \emph{User Memory} and \emph{Spark Memory}. \emph{Reserved Memory} is used for storing some internal objects of Spark, which set with a static size. Aside from this part, Spark allocates 25\% of the remaining heap memory as user Memory, this part of memory is managed by user totally. The rest of memory is managed by Spark runtime environment, which determines whether the workload is executed successful or not. This part of memory can be divided into two parts: \emph{Storage Memory} is used for caching RDD data and \emph{Execution Memory} is used for the intermediate data sorting, merging and shuffling during the data computation. Among Spark Memory, the scale and boundary of Execution and Storage part can be adjusted dynamically during the workload execution. Due to the fixed set of Reserved Memory, the main challenge of Spark workloads' memory capacity configuration lies on the configuration of Spark Memory, where most data processed and produced by the spark task are stored and most memory contention occur. Once the Storage Memory capacity is decided, the User Memory capacity can be made out according to the size ratio of 25\%:75\%. According to the analysis of Spark memory management, we can define the memory capacity requirement of Spark executor($Mem_{cap}$) as the following calculation:

\begin{equation}
Mem_{cap}=(Mem_{exe}+Mem_{sto})+UM+RM
\end{equation}
Where the $Mem_{exe}$, $Mem_{sto}$, UM and RM represent the Execution Memory, Storage Memory, User Memory and the Reservered Memory respectively.

Current approach adopted by Spark is to configure memory capacity based on user's specification. User should configure the memory capacity of each Spark executor by the parameter {\textquotedblleft}spark.executor.memory{\textquotedblright}. As the statement in[6], workloads have a tradeoff curve for memory configuration. Small configuration will cause the performance problem, while excess memory will waste memory resource. We can infer that there is a {\textquotedblleft}proper configuration{\textquotedblright} for a specific workloads, however, making such a proper configuration needs deep knowledge about their workloads and the runtime environment of the framework. \textbf{The {\textquotedblleft}proper configuration{\textquotedblright} is the optimum point by trading off the performance and the resource utilization, which can be find out manually.} Furthermore, such a {\textquotedblleft}proper configuration{\textquotedblright} is different across the diverse workloads, this is a tough challenge on memory capacity configuration.

\begin{figure}[!t]
\centerline{\includegraphics[width=2in]{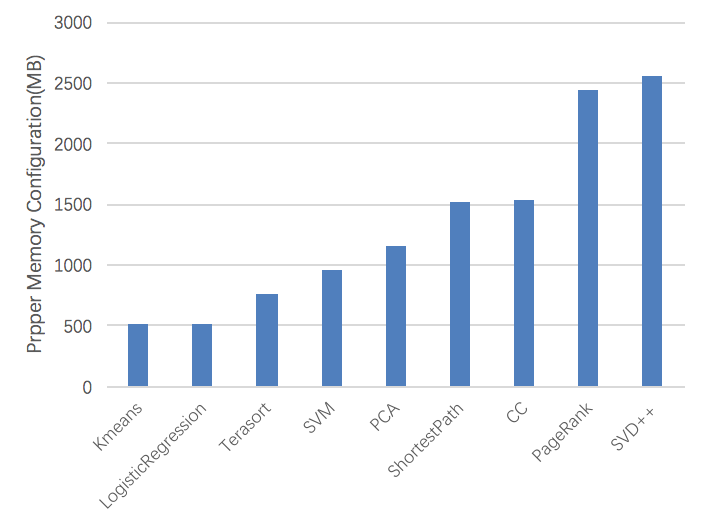}}
\caption{Memory capacity requirements among different workloads}
\label{fig_sim}
\end{figure}

\subsection{Motivation}
Our motivation is that for Spark workloads, the memory capacity requirement can be classified and evaluated. This motivation is based on two major observations of Spark workloads. First, the input data size of the specific workload has a strong relevance to its memory requirements. Second, different workloads have different memory capacity requirements and their requirements can be classified. We conduct the study under the Spark 2.1, which is the latest Spark version. For workloads, we choose SparkBench, which is a comprehensive benchmark suite for Spark. All of the experiments are done in a two node cluster. Each node has a 4-core 3.3GHz Intel Core i5-6600 processors and 4GB memory.

\begin{enumerate}
\item
\textbf{The memory capacity requirement can be evaluated, and the input data size of the workload can reflect its memory requirements.} In order to study the memory requirement of a specific workload, we measure the proper memory configuration manually. We choose four workloads to evaluate the proper memory configuration under five different input sizes.
Fig. 2 shows that the input data size of the workload has a strong relevance to its memory requirement. With the increase of the input data size,  the memory requirement grows accordingly. On the other hand, the similar growing trend is shared among the observed workloads. Hence, we can infer that the workload's memory requirements can be evaluated and quantified as the monotonic function of the input data size.

\item
\textbf{Diverse workloads have diverse memory capacity requirement and their requirements can be classified.} In order to measure the memory requirement of different workloads, we choose 9 workloads with the same input data size, 600MB, the result is shown in Fig. 3. We can see that the memory requirements among different workloads are different, for example, for the same input data size, SVD++ and PageRank require the largest memory capacity of around 2.5GB, while the KMeans and LogisticRegression only require about 512MB. Therefore, the memory requirement across different workloads are vary widly. On the other hand, when the input size is same, these workloads can fall into multiple categories according to the amount range of the memory requirement.
\end{enumerate}

\section{Methodology of WSMC}
\begin{figure}[!t]
\centerline{\includegraphics[width=3.5in]{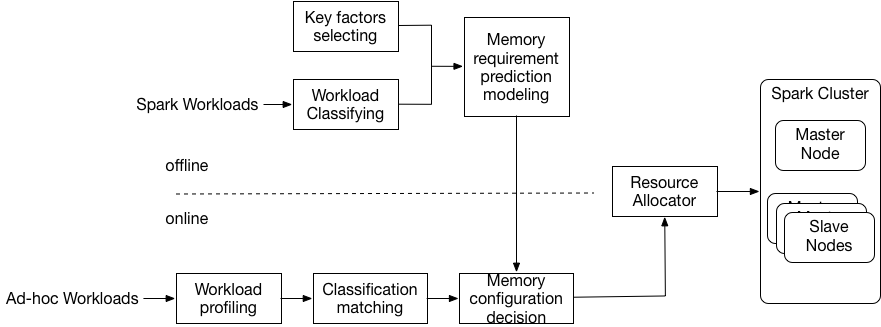}}
\caption{Overview of WSMC}
\label{fig_sim}
\end{figure}
\begin{figure*}[!t]
\centerline{\includegraphics[width=6in]{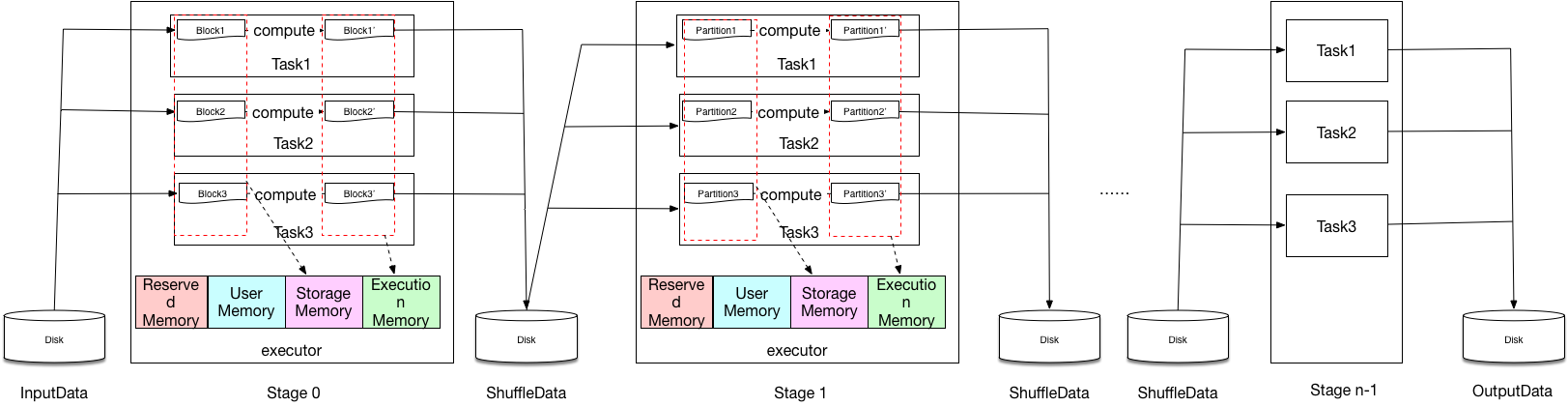}}
\caption{Data processing mechanism of Spark}
\label{fig_sim}
\end{figure*}
\subsection{Overview}

As shown in Fig. 4, the methodology of WSMC includes two phases, offline phase and online phase. In the offline phase, first, with the typical benchmark, Spark workloads are classified into multiple categories based on the diversity in their memory consumption patterns. The key factors impacting on the workload’s memory capacity requirement are also decided. Second, for each workload category, with the consideration of the key factors, the workload-specific memory requirement prediction model is established. Then in the online phase, when user submit an ad-hoc workload, the workload is profiled on its memory consumption pattern and matched into one of the above categories. The workload’s memory capacity configuration is finally calculated in accordance to the corresponding memory requirement prediction.

\subsection{Key Factors Selection}

Besides the input data size, all the key factors impacting on the Spark workload{\textquoteright}s memory requirement need to be decided as the basis of the memory capacity configuration. As shown in Fig. 5, the data processing mechanism of Spark platform can be described as follows. The Spark application is organized as the DAG-like data processing stages. Each stage can be mapped into a set of parallel tasks. For each application, Spark launches a set of JVM-based executors to execute its stages. All the stages reuse the allocated executors and execute in serial. Tasks of the first stage will load the input data into the Storage Memory of the executors before the data processing through the {\textquotedblleft}data loading{\textquotedblright} operation. Each task in the first stage need to load a data block to process. For the following stages, each task will read a partition of the result data of its parent stage through the {\textquotedblleft}shuffle read{\textquotedblright} operation and store in the Storage Memory. Each task will also cache its own result data into the Execution Memory and flush to the local disk through the {\textquotedblleft}shuffle write{\textquotedblright} operation. In this paper, for each following stage, we call the total data read in {\textquotedblleft}shuffle read{\textquotedblright} operation and written in the {\textquotedblleft}shuffle write{\textquotedblright} operation as the shuffle data. 
\begin{equation}
\begin{split}
Data_{shuf_{i}}=Data_{sr_{i}}+Data_{sw_{i}}
\end{split}
\end{equation}
Where the $Data_{sr_{i}}$ and $Data_{sw_{i}}$ represent the data size of {\textquotedblleft}shuffle read{\textquotedblright} and {\textquotedblleft}shuffle write{\textquotedblright} operation on the $i$th stage respectively, and the shuffle data size is the maximum value of $Data_{shuf_{i}}$.
\begin{equation}
Data_{shuf} = \max\{Data_{shuf_{i}}\}{i \in \{1,2,3...n\}}
\end{equation}
Where $n$ is the total number of stages. Besides, the reused input data will be cached in the memory if needed.

As mentioned before, in Spark, the memory capacity configuration is referred as the configuration of the individual executor. Besides the input data size, there are four more factors, including the shuffle data size, the data block size, the total task number of the following stages, the concurrent task number of an executor(which called task parallelism that we mentioned about) can have the impact on an executor{\textquoteright}s memory requirement. This is for that the data block size determines the data loading size of each task in the first stage. In the following stages, the shuffle data size, the total task number determine the data  size of each task{\textquoteright}s {\textquotedblleft}shuffle read{\textquotedblright} and {\textquotedblleft}shuffle write{\textquotedblright} data, while the concurrent task number of an executor determin the total data size that need to be hold on a executor. It is obviously that all these five factors can cover the main memory access behaviors of the Spark workloads.

Except of the shuffle data size, all the above five factors, can be obtained before the workload execution. However, with the knowledge of the workload’s characteristic, the shuffle data size can be deduced with the input data size. It is obvious that our model is feasible to predict the memory requirement of the Spark executor with the consideration of these five factors. 

\begin{figure}[!t]
\centerline{\includegraphics[width=2in]{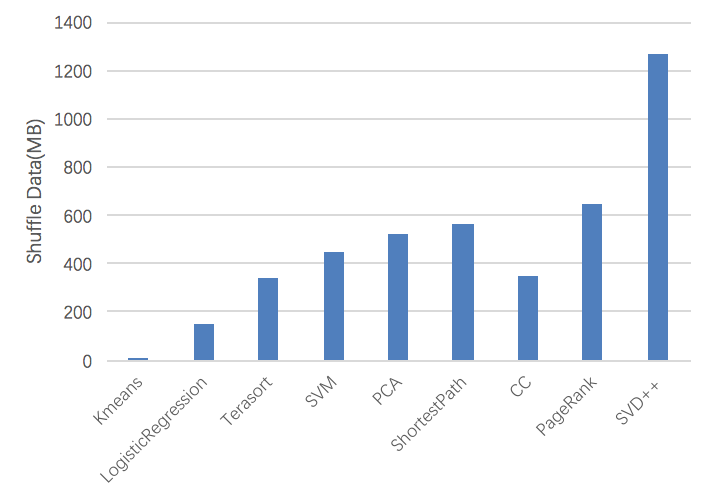}}
\caption{Shuffle data size of different workloads under same input data size}
\label{fig_sim}
\end{figure}
\subsection{Classifying Spark Workloads}
\textbf{Classification Criteria of the Workloads}
As described in Section II, with the same input data size, the amount range of memory requirement of typical Spark workloads can be classified into several categories. However, the specific memory consumption characteristic taken as the criteria of the workload classification need to be figured out. Such characteristic can not only determine the memory requirement evaluation of workload categories, but enable the ad-hoc workloads to match into those categories. 

We decide the specific characteristic with the consideration of the above five factors that impact on the workload{\textquoteright}s memory requirement. As described above, the input data size mainly determine the memory requirement of the first execution stage, while the shuffle data size determine that of the following stages. As Fig.3 shows, with the same configuration of data block size and total task number for shuffle operations, typical Spark workloads have diverse memory requirements even with the same input data size(600MB). Hence, the variety of the shuffle data size may determine the workload classification. To verify it, we examine the shuffle data size of workloads demonstrated in Fig.3. As shown in Fig.6, the workloads can also be classified according to the size range of the shuffle data. For the PageRank and SVD++, the shuffle data size is larger than the input data size. The shuffle data size of SVD++ and PageRank are over than 600MB, especially the shuffle data size of SVD++ is over than 1200MB. However, for the LogisticRegression and KMeans, their shuffle data size is less than half of their input data size. For all the other workloads, the shuffle data size is as much as 0.5 to 1 time of the input data size. More interesting, the resulting workload categories are quite same as those with the memory requirement classification. We therefore determine that the ratio of the shuffle data size to the input data size can be the specific memory consumption characteristic to classify Spark workloads and define it as {\textquotedblleft}Data Expansion Ratio{\textquotedblright}. The Data Expansion Ratio of the Spark workload can be calculated as follows:
\begin{equation}
\alpha=\frac{Data_{shuf}}{Data_{input}}
\end{equation}
Where the $Data_{input}$ and $Data_{shuf}$ represent a specific input data size and the corresponding shuffle data size respectively.

\textbf{Methodology of Classifying Spark Workloads}
We describe the methodology of classifying Spark workloads with the metric of Data Expansion Ratio.

First, the Spark workload is executed with a set of input data in turn. The input data set can be expressed as $DS$, $DS$ = $\{ds_1, ds_2, ..., ds_n\}$, which is sorted with the data size in ascending order. For each input data $ds_i$ the Data Expansion Ratio $\alpha_{i}$ can be calculated using formula(4). For the sake of accuracy, the Data Expansion Ratio of the spark workload is expressed as the average of all the Data Expansion Ratio.

Second, as shown in Table I, the spark workloads can be classified into three categories according to the value of their Data Expansion Ratio.

\begin{enumerate}
\item
\textbf{Expanding}: Workloads that generate much larger volume of the shuffle data than that of their input data.
\item
\textbf{Shrinking}: Workloads that generate much smaller volume of the shuffle data than that of their input data, generally less than half of the input data. 
\item
\textbf{Medium}: The volume of shuffle data is between the above two categories.
\end{enumerate}

\begin{table}
\caption{Classification of the workload based on Data Expansion Ratio}
\centering
\begin{tabular}{|l|l|}
\hline\noalign{\smallskip}
Category Name & Classification Condition \\
\noalign{\smallskip}
\hline
\noalign{\smallskip}
Expanding& $\alpha \geq  1$ \\
Shrinking& $\alpha \leq  0.5$ \\
Medium& $0.5 < \alpha < 1 $ \\
\hline
\end{tabular}
\end{table}

Finally, as the Expanding category can cover a  very   large   range of the Data Expansion Ratio, which is from 1 to infinity in theory, we choose the {\textquotedblleft}Increasing Rate of Data Expansion{\textquotedblright} to do the further division. The Increasing Rate of Data Expansion of the spark workload is examined based on the input data set $DS$ and can be defined as follows:

\begin{equation}
inc_{shuf_i}=\frac{Data_{shuf_{i+1}}-Data_{shuf_i}}{Data_{input_{i+1}}-Data_{input_i}}(i\in{1,2,...n-1})
\end{equation}
And we calculate the average of $inc_{shuf_i}$ as the value of $inc_{shuf}$. 

As shown in Table II, the expanding category can be further divided into two categories according to the workload’s Increasing Rate of Data Expansion.

\begin{enumerate}
\item
\textbf{Expanding.Rapid}: Expanding workloads with the Increasing Rate of Data Expanion greater than 2.
\item
\textbf{Expanding.Medium}: Expanding workloads with the Increasing Rate of Data Expanion less than 2.
\end{enumerate}

\begin{table}
\caption{Classification of the workload based on Increasing Rate of Data Expanion}
\centering
\begin{tabular}{|l|l|}
\hline\noalign{\smallskip}
Category Name & Classification Condition \\
\noalign{\smallskip}
\hline
\noalign{\smallskip}
Expanding.Rapid & $inc_{shuf} \geq  2$ \\
Expanding.Medium & $ inc_{shuf} < 2 $ \\
\hline
\end{tabular}
\end{table}

\begin{table}
\caption{The values of Data Expansion Factor}
\centering
\begin{tabular}{|l|l|}
\hline\noalign{\smallskip}
Class Name & $factor_{shuf}$ \\
\noalign{\smallskip}
\hline
\noalign{\smallskip}
Expanding.Rapid & 4(3+1) \\
Expanding.Medium & 3(3+0) \\
Medium & 2 \\
Shrinking & 1 \\
\hline
\end{tabular}
\end{table}
Base on the analysis above, we can divide the Spark workload into four categories, including {\textquotedblleft}Expanding.Rapid{\textquotedblright}, {\textquotedblleft}Expanding.Medium{\textquotedblright}, {\textquotedblleft}Medium{\textquotedblright} and {\textquotedblleft}Shrinking{\textquotedblright}.

\subsection{Prediction Model of Memory Requirement of Spark Workloads}

Due to that the memory capacity configuration in Spark is applied to each individual executor, we predict the maximum amount of memory that an executor requires among multiple processing stages and take it as the required memory capacity of a Spark workload. We establish the prediction model with the consideration of the four relevant factors described above. All the factors, except the shuffle data size, can be obtained via the configuration parameters. As mentioned above, the shuffle data size can be expressed as the expansion of the input data size. We predict the shuffle data size in a workload-specific way as follows:
\begin{equation}
Data_{shuf}=factor_{shuf} \times Data_{input}
\end{equation}
Where $Data_{input}$ is the input data size, $factor_{shuf}$ is the expansion factor. We evaluated nine workloads of SparkBench, and get the experiment value of $factor_{shuf}$. The detail $factor_{shuf}$ setting of different class is shown as Table \uppercase\expandafter{\romannumeral3}.

On the first stage, we mainly focus on the {\textquotedblleft}data loading{\textquotedblright} operation. Hence, we define the maximum amount of memory required by an executor on the first processing stage as follows:
\begin{equation}
Mem_{s_f} = Size_{block}\times min\{Task_{ex}, \frac{Data_{input}}{Size_{block}}\}
\end{equation}
Where the $Size_{block}$ represent the block size, the $Task_{ex}$ represent the maximum concurrent task number of an executor, which can be set by the parameter {\textquotedblleft}spark.executor.cores{\textquotedblright}. The $\frac{Data_{input}}{Size_{block}}$ represent the total task number of first stage.

On the other stages we focus on the {\textquotedblleft}shuffle read{\textquotedblright} and {\textquotedblleft}shuffle write{\textquotedblright} operations, the maximum amount of memory required can be calculated as follows:

\begin{tiny}
\begin{equation}
\begin{split}
Mem_{s_o}=\frac{Data_{input}} {Num_{ex}} \times \beta +\frac{Data_{shuf}}{parallelism} \times min\{Task_{ex}, parallelism\}
\end{split}
\end{equation}
\end{tiny}
Where the $parallelism$ represent the total task number for shuffle operations, which can be set by the parameter {\textquotedblleft}spark.default.parallelism{\textquotedblright}. The $\beta$ represent the condition whether the input data need be cached in the executor, set 1 as need to be cached in the executor else set 0. The ${Num_{ex}}$ is the number of executor, which can be calculated by the input data size, the the concurrent task number of an executor and the block data size. The calculation is shown as follows:
\begin{equation}
Num_{ex} = \lceil{\frac{Data_{input}}{Task_{ex}\times Size_{block}}}\rceil
\end{equation}
Therefore, the memory capacity of an executor is the maximum of the all stages.
\begin{equation}
Mem_{spark}=max\{Mem_{s_f}, Mem_{s_o}\}
\end{equation}

According to the Spark memory management in Section 2. The calculation of Spark executor's memory capacity configuration of each executor is calculated as follows:
\begin{equation}
\begin{split}
Mem_{cap}=Mem_{spark}\times \frac{4}{3}+RM
\end{split}
\end{equation}
where the $RM$ is represented \emph{Reserved Memory}, $RM$ is a static value(300MB in default).

All the fomulas of our prediction model can be represented by the key factors that we listed above, which are retrievable. Hence, our prediction model is feasible.

\subsection{Online Workload Profiling and Classification Matching}
A Spark environment needs to run the ad-hoc workloads. We cannot collect the shuffle data until it has been executed. This makes us difficult to make a proper memory configuration. WSMC addressed this challange by worloads profiling. We running the ad-hoc workload in a set of small-sized input data. For example, for a specific workload $WL$, first, we execute $WL$ in 10MB, 20MB, 30MB, 40MB, 50MB respectively, in corresponding, the shuffle data size are 20MB, 40MB, 60MB, 80MB and 100MB. Second, we calulate the Data Expanding Ratio of $WL$, the $\alpha$ of $WL$ is 2, which belongs to the 
{\textquotedblleft}Expanding{\textquotedblright} workloads. Hence, we calculate the Increasing Rate of Data Expansion, the $ratio_{shuf}$ of $WL$ is also equals 2. Therefore, $WL$ is a {\textquotedblleft}Expanding.Rapid{\textquotedblright} workload. According to the Table \uppercase\expandafter{\romannumeral3}, the $factor_{shuf}$ is 4. We assume that the $Size_{block}$ is 128MB the $Task_{ex}$ and $parallelism$ are both 1, the $\beta$ is set as 1, when the input data size is 1GB, the memory capasity of $WL$ is 1153MB.

\subsection{Disccusion}
Through the description above, we propose a workload-specific methodology to predict the memory capacity configuration of Spark workload, including the classification methodology and the prediction model of Spark workloads' memory capacity. However, there are some conditions that our study not take into account, our prediction model assume that the input data size is lower than the total memory capacity of the cluster, this is due to user adopt the in-memory framework for taking most use of memory to improve the performance of their workloads. When the input data size is over than the maximum memory capacity of cluster, frquent data transformation between disk and memory will cause significant performance decline. Thus, our work not take this situation into account, while we will consider such sitiation in our future work.

\begin{table*}
\newcommand{\tabincell}[2]{\begin{tabular}{@{}#1@{}}#2\end{tabular}}
\caption{Workloads' memory capacity configurations guided by WSMC}
\centering
\begin{tabular}{lllcl}
\hline\noalign{\smallskip}
No & Name & Input Size(MB) & Category & \tabincell{c} {Memory \\ Configuration(MB)}\\
\noalign{\smallskip}
\hline
\noalign{\smallskip}
1 & SVD++ & 100 200 300 & Expanding.Rapid & 860 1160 1460\\
2 & PageRank & 100 200 300 & Expanding.Rapid & 860 1160 1460\\
3 & TriangleCount & 100 200 300 & Expanding.Rapid & 860 1160 1460\\
4 & ShortestPath & 400 500 600 & Expanding.Medium & 1040 1200 1360 \\
5 & Terasort & \tabincell{l} {600 700 800} & Medium & \tabincell{l}{1030 1130 1230}\\
6 & ConnectedComponent & \tabincell{l} {600 700 800} & Medium & \tabincell{l}{1030 1130 1230}\\
7 & SVM & \tabincell{l} {600 700 800} & Medium & \tabincell{l}{1030 1130 1230}\\
8 & PCA & \tabincell{l} {600 700 800} & Medium & \tabincell{l}{1030 1130 1230}\\
9 & DecisionTree & \tabincell{l} {600 700 800} & Medium & \tabincell{l} {1030 1130 1230}\\
10 & KMeans &  \tabincell{l} {1200 1300 1400} & Shrinking & \tabincell{l} {1040 1090 1140}\\
11 & LogisticRegression & \tabincell{l} {1200 1300 1400} & Shrinking & \tabincell{l} {1040 1090 1140}\\
12 & Matrix Factorization & \tabincell{l} {1200 1300 1400} & Shrinking & \tabincell{l} {1040 1090 1140}\\
\hline
\end{tabular}
\end{table*}

\section{Performance Evaluation}
In this section, we conduct experiments to evaluate the efficiency of the WSMC approach. First, we select twelve workloads from SparkBench as the workloads of experiments. Second, we choose the {\textquotedblleft}proper{\textquotedblright} configuration and the default configuration as the comparisons.

\subsection{Experimental Methodology}
We select twelve workloads from SparkBench as the workloads of experiments. Table \uppercase\expandafter{\romannumeral4} is the workloads' memory capacity configurations which guided by WSMC, we use five small-sized input data size to profile the workloads, which is 50MB, 100MB, 150MB, 200MB and 250MB respectively. As shown on the Table \uppercase\expandafter{\romannumeral4}, we evaluate each workload with three different input data size, and the corresponding memory capacity configurations are shown on the last column. In our cluster, the $Task_{ex}$ and the $parallelism$ are both 4, the $\beta$ is 1, and the $Size_{block}$ is 128MB. We execute all the workloads under three classes of configurations. The first one is the configuration under the guide of WSMC, the second one is the default configuration of Spark platform which is static configured as 2GB, the last one is the proper configuration, we get the proper configuration by trading off the performance and memory capacity, we increase the memory capacity of executor with fixed step until the curve of the performance and the memory capacity.

In order to evaluate the efficiency of the WSMC approach, we choose two metrics which is the execution time of the workload and the memory usage of the system.

\subsection{Execution Time}

Fig. 7 demonstrates the normalized execution time of workloads under different memory configurations. in the Fig. 8, we normalized the execution time of workloads under different memory configurations, and set the execution time of default configuration as the normalized object.

From Fig. 7, we can find that the execution time with the configuration under the guide WSMC and the proper configuration are slightly larger than that with default configuration of Spark. Compare to the default configuration of Spark, the execution time increase 4.68\% and 4.83\% respectively. However, their gap is very small.

On the one hand, comparing WSMC with default configuration, for different categories of workloads, we can see that the execution time of  {\textquotedblleft}Expanding{\textquotedblright} workloads increase about 5.5\% on average, the execution time of {\textquotedblleft}Medium{\textquotedblright} workloads increase about 4.5\% on average, and the execution time of {\textquotedblleft}Shrinking{\textquotedblright} workloads increase 3.7\% on average. We can also find that the performance penalty of workloads which belong to the same category are different, due to their different intermediate data size. For example, SVD++, TriangleCount and PageRank are belong to the {\textquotedblleft} Expanding, Rapid{\textquotedblright} application, while the average $\alpha$ of SVD++ is over than 10, but that of PageRank is about 2. Due to our approach only takes the main part of memory consuming, more intermediate data is more likely cause the memory contension, which can bring extra performance penalty. However, we can find that the performance penalty are quite slight, and with the execution time increased 6.25\% by maximum.

On the other hand, comparing WSMC with the proper configuration, the execution time with the configuration under the guide of WSMC is very close. The performance gap is about 1\%.
\begin{figure}[!t]
\centerline{\includegraphics[width=2in]{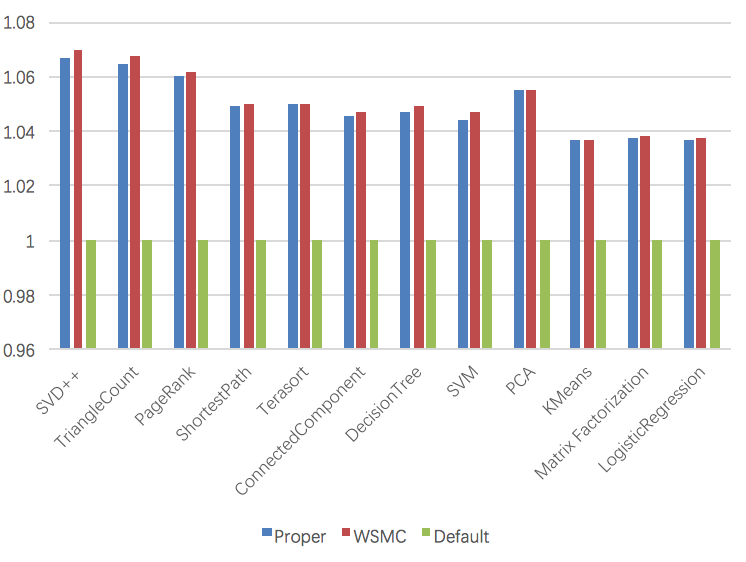}}
\caption{The normalized execution time of different memory configurations}
\label{fig_sim}
\end{figure}
\subsection{The Memory Usage}
\begin{figure}[!t]
\centerline{\includegraphics[width=2in]{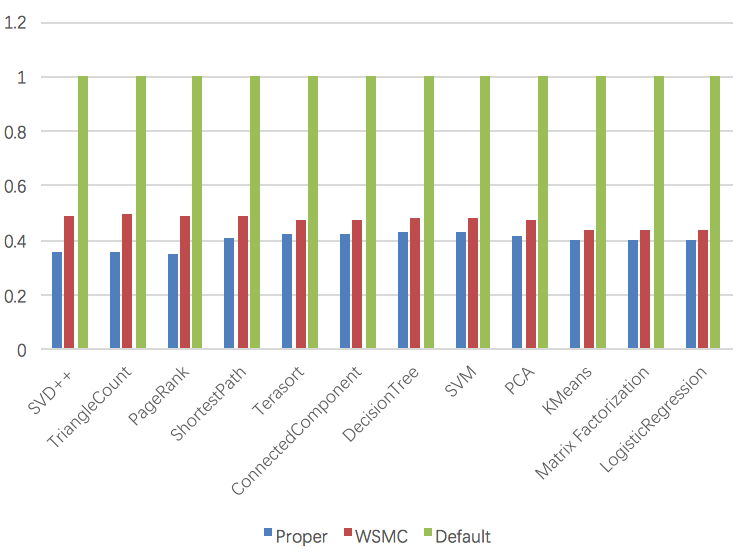}}
\caption{The normalized memory usage under different memory configurations}
\label{fig_sim}
\end{figure}

Fig.8 demonstrates the normalized memory usage of workloads under different memory configurations. in the Fig.8, we normalized the memory usage of workloads under different memory configurations, and set the memory usage of default configuration as the normalized object.

As shown on the Fig. 8, compared to the default configuration, the configuration with WSMC can reduce the memory usage of the system significantly, and the minimal value is 36.5\% and the maximum one is 58\%. Lacking of deep knowledge of the Spark mechanism, non-expert users always adopt the excessive for ensuring the performance of their workloads, and the configuration always from the official document of Spark. From the experiments, we can see that using WSMC can reduce the memory usage of the system significantly. Furthermore, compare to the default configuration of Spark, the {\textquotedblleft}Expanding{\textquotedblright} applications can saves over than 50\% of memory, the {\textquotedblleft}Medium{\textquotedblright} application can save 40\% of memory, and the {\textquotedblleft}Shrinking{\textquotedblright} applications save 36.5\% of memory. Though the classification of workloads and the observation of the trend of memory requirement under different input data size.

On the other hand, compare to the proper configuration, the configuration under the guide of WSMC uses about 7\% of memory more than proper configuration. Furthermore, the {\textquotedblleft}Expanding{\textquotedblright} use 14\% of memory more than the default memory configuration, and the the {\textquotedblleft}Medium{\textquotedblright} uses about 5\% memory more than the default memory configuration, and the {\textquotedblleft}Shrinking{\textquotedblright} use 3\% memory more than the default configuration. From the result, we can find that the configuration under the guide of WSMC is more closely to the proper memory configuration than the default memory configuration of Spark.

\subsection{Summary}
Through our experiment, we prove that the memory requirement of Spark workloads can be evaluated and classified. According to the classification, we can predicted the memory capacity requirement of ad-hoc Spark workloads accurately. The memory configuration under the guide of WSMC only about 7\% memory wasted than that of the proper configuration.

Besides, WSMC model can reduce the memory usage significantly with slight performance penalty. Compare to the default configuration of Spark, the configuration based on WSMC can reduce the memory usage by 40\%, however the augmentation of execution time is only about 5\%.

So, we can find that the WSMC can guide the memory capacity configurations of Spark efficiently.

\section{Related Work}
The memory management for in-memory data analytic platforms is a hot topic.
MEMRUNE[10] is one of the dynamic memory management system for in-memory platform, which can adjust the space of caching and execution space dynamically and provide policies of data eviction for caching data. Besides, after Spark version 1.5, Spark community proposed a project named {\textquotedblleft}Tunsten{\textquotedblright}, the motivation of this project is to take the most use of the resource. For memory management, this project allows the Spark platform to manage off-heap memory directly, which can reduce the GC opeation and improve the performance of Spark workloads. Besides, this project also provide a cache-aware structure to process and store data efficiency. However, these works focus on how to  managing the memory space more effeciently, rather than memory capacity planning, our work and those works are complementary.

In addtion to the memory management, some researches are focus on the configuration optimization in order to tunning the performance of workloads. Some works provide some suggestions of configuration optimization by a comprehensive performance evaluation of target platforms. For example, Pan[9] et al analyze the I/O characteristic on Hadoop platform, they choose four typical big data workloads from BigdataBench[13] and measure the impact on the I/O characteristic through changing the key parameters, finally give some suggestions for improving the performance. The limitation of these works is that they cannot give a proper configuration to a ad-hoc workload. Some works aware the limitation, propose some online approaches to optimize the configuration. For instance, AROMA[14] and MRONLINE[12] is such system for Hadoop platform. Both of these two systems will first collect some performance data of workload that have been executed and construct a knowledge base. When user submit an ad-hoc workload, they will run these workloads in a small-sized data size or cluster with default configuration, collect their performance data, match them with the offline knowledge base, and the system will provide a proper configuration for this ad-hoc workload automatically. Our work learn from these approaches and applying them into multi-stage in-memory data analysis platform.

Besides, there are lots of works are focus on the optimization of Spark platform. Such as programming models[4] and APIs, on the other hand, some new hardware resources are integrated into the Spark platform, such as GPUs and FPGAs, user can utilize these resources by using the extended APIs to optimize their workloads, the module of resource management will manage these resources. At the same time, Spark is adopted in more and more domains, such as DNA Analyzing and Wether Forcasting, these works makes Spark plays more and more important role on data analysing.

\section{Conclusion}
In this paper, We propose WSMC, a workload-specific memory capacity configuration approach for spark workloads. In order to classifing the workloads, we take the Data Expansion Ratio, which is defined as the maximum ratio of the shuffle data to the input data among the multiple processing stages of a Spark workload,as the basic metric for the workload classification. Furthermore, for each workload category, establish a workload-specific memory requirement prediction model for Spark workloads with the parameters of the input data size, the shuffle data size, the task parallelism and the data block size. Through the comprehensive evaluations with SparkBench workloads, we found that contrasting with the defaultconfigurations, the configuration with the guide of WSMC can save over 40\% memory resource with a slight workload performance degradation (only 5\%) and compare to the proper configuration, which is found out manually, the configuration with the guide of WSMC leads to only 7\% increase in the memory waste with the workload's performance slight improvement (about 1\%).

% or


% \appendices

% \section{}

% \section*{Acknowledgment}


\ifCLASSOPTIONcaptionsoff
  \newpage
\fi



%\bibliographystyle{IEEEtran}

\section{Acknowledgements}
The work was partially supported by National Key R\&D Program of China under grant number 2017YFC0803300, Beijing Natural Science Foundation under Grant 4172004, National Natural Science Foundation of China under grant number 91546111, Beijing Municipal Education Commission Science and Technology Program under grant number KZ201610005009 and KM201610005022.


% \end{IEEEbiography}


% % insert where needed to balance the two columns on the last page with
% %\newpage



%\vfill


\end{document}